\documentstyle[preprint,prb,aps]{revtex}
\newcommand{\bmp}{{\mbox{\boldmath $p$}}}

\newcommand{\bmr}{{\mbox{\boldmath $r$}}}

\newcommand{\bmq}{{\mbox{\boldmath $q$}}}

\newcommand{\AAr}{{\rm \AA}}
\newcommand{\Ain}{{\rm {\AA}^{-1}}}
\newcommand{\ppp}{{\frac {d\bmp}{(2\pi)^3} }}

\begin{document}
\preprint {WIS-97/01/Jan-PH}
\draft
 
\date{\today}
\title{Description of recent
large-$q$ neutron inclusive scattering data from liquid $^4$He}
\author{A.S. Rinat and M.F. Taragin}
\address{Department of Particle Physics, Weizmann Institute of Science,
         Rehovot 76100, Israel}
\author{F.Mazzanti and A. Polls}
\address{Department d'Estructura e Constituents de la Materia,
         Diagonal 645, Universitat de Barcelona,
         E-08028 Barcelona, Spain}
\maketitle
\begin{abstract}
 
We report dynamical  calculations for large-$q$ structure functions of
liquid $^4$He  at $T$=1.6 and  2.3 K and  compare those with  recent
MARI data.  We extend those calculations far beyond the experimental
range $q\le 29\Ain$ in  order to study the  approach of the
response to its asymptotic limit for  a system with interactions having a
strong  short-range   repulsion.  We find only small deviations from
theoretical $1/q$ behavior, valid for smooth $V$. We repeat an
extraction by Glyde et al of cumulant coefficients from data.
We argue that fits determine the single
atom momentum distribution, but express doubt as to the extraction of
meaningful Final State Interaction parameters.
 
\end{abstract}
\pacs{}
 
\section{Introduction}
 
In  the following  we discuss different aspects of  the response  of
liquid $^4$He.   First we  report predictions which  are compared  with
most  recent data.   Next, we compute the response
for  $q\lesssim 300 \Ain$ in order to study how
Final State  Interaction (FSI) effects   vanish for those large $q$.
In the  end we present results of a new
model-independent cumulant
analysis  of data in  order to extract  the single-atom
momentum distribution and interaction parameters.

We address below recent precision data for temperatures
below and above  the transition temperature $T_c$, which  have been taken
at the Rutherford ISIS facility by means of the MARI spectrometer.  Those
by Andersen  $et\,al$ span neutron momentum  transfers $3\leq q(\Ain)\leq
10$  for  $T$=1.42  K  and $3\leq  q(\Ain)\leq  17$  for  $T$=2.5  K
\cite{and1}, while Azuah's measurements covered $10 \leq q (\Ain)\leq 29$
for $T=$1.6, 2.3 \cite{azuah}.  The  above expand  in scope  previous
information taken  a few years  ago at the  IPNS facility at  Argonne for
$q\le 23.1\Ain$ \cite{sos}.
 
From inclusive cross sections one extracts the linear response. The
latter is a  function of  two parameters  $q$  and  $\omega$,  which in
the scattering experiment are  the momentum  and
energy, transferred  from the projectile  to the target.  For medium
and large $q$ those responses
contain  information   on  the  target,  such   as  the  momentum
distribution of  the constituents and prescribed  manifestations of their
interaction.
 
A dynamical calculation  of
the response requires as  input the atom-atom
interaction and  groundstate information, which for  the above $q$-regime
are  primarily  the  single-atom  momentum distribution  $n(p)$  and  the
semi-diagonal two-particle density matrix.
 
Using variations  of much  the same  theory, calculations have been made
before for   medium-$q$, as well as for the higher-$q$
Argonne data \cite{sil,rt,rt1,cako,fer}.
To  our  knowledge no ab initio calculations of the MARI data have been
performed until now.
 
The same MARI data  have recently been approached in an entirely
different fashion with the purpose
to determine in a model-independent way the
dominant coefficients  in the cumulant  expansions of the  asymptotic and
FSI parts of the response \cite{glyde,glyde1}.
Good fits  to the data were obtained, but those have little in common
with dynamic calculations. The latter  use as $input$
$n(p)$,  additional groundstate information  and $V$ whereas, ideally,
from  cumulant fits  $n(p)$ and properties of $V$ are $extracted$.
 
As a major result of the above analysis, Glyde $et\,al$ report the
reconstruction of  the single-atom  momentum distribution $n(p)$  in good
agreement with accurate theoretical predictions \cite{cepo,wp}.  However,
a less satisfactory feature is the extracted  dominant FSI
cumulant coefficient
which, dependent on the analysis, is reported to be less than 0.65
times  the calculated value. One then wonders whether the apparent
partial fit may have consequences
on the precision of the reconstructed $n(p)$.
 
The following program emerges from the above observations.  In Section II
we  outline an  approach to  high-$q$  responses.
In Section III  we report computations of the high-$q$ measurements
using the MARI spectrometer
and compare those predictions with the data.
In addition we interpret responses
computed out to very high  $q\le 300 \Ain$. The
results enable the study  of the approach of the response
to its asymptotic  limit for systems with a  strong short-range repulsion
in the  interaction between the  constituents. In Section  IV we present
fits for cumulant parameters for $T$=2.3K
and compare those with  similar results by Glyde
$et\,al$\cite{glyde,glyde1,ken}.
We discuss the discrepancy between the
calculated and the extracted  FSI
parameters and attribute it to the truncation of the cumulant series.
In the conclusion we estimate that both experimental and
theoretical studies of the response of liquid $^4$He at high $q$
may have reached a  degree of sophistication, beyond which there is
little prospect to gain new information.
 
 \section {Descriptions of the linear response for high q.}
 
Consider for infinitely extended liquid $^4$He the response per atom in
the form
 
\begin{eqnarray}
S(q,\omega)&=&
A^{-1}(2\pi)^{-1}\int_{-\infty}^{\infty} dt
e^{i\omega t}\langle 0|\rho_q^{\dagger}(t)\rho_q(0)|0\rangle
\nonumber\\
\phi(q,y)&=&(q/M)S(q,\omega),
\label{a1}
\end{eqnarray}
with $M$ the mass of a $^4$He atom. $\rho_q(t)$ above is the
density operator \begin{eqnarray}
\rho_q(t)&=&e^{-iHt}\rho_q(0)e^{iHt}
\nonumber\\
\rho_q(0)&=&\sum_j e^{i\bmq\bmr_j(0)}
\label{a2}
\end{eqnarray}
Strictly speaking, the symbol $\langle 0\vert...\vert 0\rangle$
should stand for a canonical average at given $T$, but we shall use
instead  the ground state in conjunction  with $T$-dependent quantities.
 
In the last line of (1) we introduce the reduced response $\phi(q,y)$
with  the energy loss $\omega$, replaced by  an  alternative
kinematic  variable $y=y(q,\omega)$ \cite{grs1,west}
\begin{eqnarray}
y=\frac {M}{q}\bigg (\omega-\frac{q^2}{2M}\bigg )
\label{a3}
\end{eqnarray}
 
Upon substitution of  (2) into (1) one generates two components
of the response. In the  incoherent part, one tracks
the same particle when propagating in the medium, while in the
coherent part one transfers momentum and energy to a particle
distinct from
the struck one. For $q\gtrsim 8 \Ain$ the
response is dominated by the incoherent part and
the coherent part one can be safely disregarded.
 
For the description of the large-$q$ response we shall exploit the theory
of Gersch, Rodriguez and Smith (GRS) for smooth interactions $V$
which leads to the following expansion for the
reduced response  in inverse powers of $q$
or of the recoil velocity $v_q=q/M$ \cite{grs1}
(We use units $\hbar=c=1$ causing
all quantities to have dimensions of  powers of $\AAr$ or $\AAr^{-1}$)
\begin{mathletters}
\label{a4}
\begin{eqnarray}
\phi(q,y)&=&\sum_{n=0}^{\infty}\bigg (\frac {1}{v_q}\bigg )^n F_n(y)
\label{a4a}\\
F_0(y)&=&{\lim_{q\to\infty}}\phi(q,y)=
(4\pi^2)^{-1} \int^{\infty}_{|y|}dp p n(p)
\label{a4b}\\
\frac{1}{v_q}F_1(y)&=&i(2\pi\rho)^{-1}\int_{-\infty}
^{\infty}\,\,ds\,e^{iys}
\int d\bmr \rho_2(\bmr,0;\bmr-s\hat\bmq,0)\tilde\chi(q;\bmr,s)
\label{a4c}\\
\tilde\chi(q,\bmr,s)
&=&-\frac {1}{v_q}\bigg \lbrack
\int_0^s ds' V(\bmr-s'\hat\bmq)-sV(\bmr-s\hat\bmq)\bigg \rbrack
\label{a4d}\\
&&...\,\, {\rm etc}.
\nonumber
\end{eqnarray}
\end{mathletters}
The function $\tilde\chi$ in (\ref{a4d}) resembles an
eikonal phase. It differs from it
because the integration limits on the line integral over the first
component of $V$ are not $(-\infty,\infty)$, as is appropriate
for on-shell scattering. The
finite limits correspond to off-shell scattering described in  the
coordinate representation. Moreover a second  interaction is implicit in
(\ref{a4d}). In the following we shall allude to the total expression
(\ref{a4d}) as the generalized eikonal phase.
 
We recall the interpretation of the lowest order terms.
For sufficiently large momentum transfer $\bmq$,
an atom  with initial momentum $\bmp$ recoils with
$p'=|\bmp+\bmq|\approx q \gg {\langle p^2\rangle}^{1/2}$, which
is larger than the average momentum of an atom in the medium and is
moreover in excess of any inverse length in  the system. The recoiling
atom moves therefore too fast to be affected by atom-atom  collisions
and the response is the asymptotic limit  $F_0(y)$
for $q,\omega  \to\infty$ at  fixed  $y$.  Eq. (\ref{a4b}) shows
its expression  in terms of the single-atom
momentum distribution, normalized as $\int d\bmp/(2\pi)^3n(p)=1$.
 
Although the GRS  theory is not a perturbation theory  in the
interaction $V$, the second term in the  series (\ref{a4a}), linear
in $V$, is entirely  due to binary
collisions (BC)  between the hit  and any other  atom.
It accounts  for the dominant  FSI
collecting all contributions $\propto 1/q$.
This is achieved  at the price of
introducing the   semi-diagonal  two-particle  density  matrix
$\rho_2$ in (\ref{a4c}).
 
In another publication Gersch and co-workers suggested an alternative
representation for the reduced response \cite{grs2}
\begin{mathletters}
\label{a5}
\begin{eqnarray}
\phi(q,y)&=&
\int dy' F_0(y-y')R(q,y')= \int \ppp n(p)R(q,y-p_z)
\label{a5a}\\
{\tilde \phi}(q,s)&=&\int_{-\infty}^{\infty}\,dy\,e^{-iys}\phi(q,y)
=\sum_n\bigg (\frac {1}{v_q}\bigg )^n \tilde F_n(s)
\nonumber\\
&=&{\tilde F}_0(s){\tilde R}(q,s)\equiv
{\tilde F}_0(s){\rm exp}[\tilde {\Omega}(q,s)]
\label{a5b}
\end{eqnarray}
\end{mathletters}
In Eq. (\ref{a5a}) the response is written
as a convolution  of its  asymptotic limit and a
FSI factor $R(q,y)$. It is frequently convenient
to use
Fourier transforms $\tilde F_i(q,s), \tilde R(q,s)...$. In particular
for the first two terms in (5b)
one has (cf. (\ref{a4b})-(\ref{a4c}))
\begin{mathletters}
\label{a6}
\begin{eqnarray}
\tilde F_0(s)&=&
\frac {\rho_1(s,0)}{\rho}=\int \frac {d\bmp}{(2\pi)^3}
e^{-i\bmp\hat\bmq s} n(p)
\label{a6a}\\
\tilde F_1(s)&=&\frac {i}{\rho}
\int d\bmr \rho_2(\bmr-s\hat\bmq,0;\bmr,0)\tilde\chi(q;\bmr,s),
\label{a6b}
\end{eqnarray}
\end{mathletters}
\noindent
with  $\rho_1(s,0)=\rho_1(\bmr-s\hat\bmq,\bmr)$, the single-atom density
matrix and $\rho=\rho_1(\bmr,\bmr)$, the number density.
 
We shall restrict ourselves below to various descriptions of
FSI due to BC, starting from the corresponding cumulant form
(\ref{a5b}) and using (\ref{a6a}) \cite{grs2}
\begin{eqnarray}
\tilde \phi(q,s)&=&
\frac {\rho_1(s,0)}{\rho}\tilde R_2(q,s)= \frac {\rho_1(s,0)}{\rho}
{\rm exp}[{\tilde \Omega_2(q,s)}]
\nonumber\\
\tilde\Omega_2(q,s)&=&
i\rho\int d\bmr \zeta_2(\bmr,s){\omega_2}(q,\bmr,s),
\label{a7}
\end{eqnarray}
\noindent
where $\tilde R_2(q,s)$ and $\tilde \Omega_2(q,s)$ are the BC
approximation to the corresponding quantities defined in (\ref{a5a}) and
(\ref{a5b}). $\zeta_2$ above is  defined by
\begin{mathletters}
\label{a8}
\begin{eqnarray}
\zeta_2(\bmr,s)&=&
\frac{\rho_2(\bmr-s\hat{\bmq},0;\bmr,0)}{\rho\rho_1(s,0)}
\label{a8a}\\
\zeta_2(\bmr,0)&=&g_2(\bmr)
\label{a8b}
\end{eqnarray}
\end{mathletters}
with $g_2$  the pair-distribution function.
 
Eq. (\ref{a7}) is the most general cumulant form in the BC approximation
for the FSI phase $\tilde\Omega_2(q,s)={\rm ln}[\tilde R_2(q,s)$],
and distinguishes through
$\omega_2$ between different dynamical
approaches. For instance for smooth interactions $V$, which would
allow for an
expansion
of the exponential in (\ref{a7}), comparison of (\ref{a7}), (\ref{a5b})
and (\ref{a4d}) shows $\omega_2$ to be the generalized eikonal phase
\begin{eqnarray}
{\omega}_{2,V}(q,\bmr,s)=\tilde\chi(q,\bmr,s)
\label{a9}
\end{eqnarray}
 
For interactions with a strong short-range
repulsion, the line
integral over  $V$ in the (off-shell) phase
(\ref{a4d}) which enters the dominant BC FSI
contribution $\tilde F_1(s)$, Eq. (\ref{a6b}), may
produce large and even divergent integrals. The standard method
to tackle those difficulties
is by partial summation of  selected higher order  terms
\begin{eqnarray}
i\omega_{2,V}\to i\omega_{2,t}=e^{i\tilde \chi}-1,
\label{a10}
\end{eqnarray}
which amounts to replacing the $'$bare$'$ $V$ by a
$q$-dependent effective interaction
$V\to V_{eff}(q)=\tilde t(q)$, the latter being the
off-shell $t$-matrix, in turn generated by $V$. Moreover the
propagation in between collisions is described in the
eikonal approximation \cite{rt1,bespro}.
 
In an alternative regularization for an  atom-atom interaction with
a strong short-range repulsion, one replaces
the generalized eikonal phase (\ref{a4d})  by
a semi-classical approximation\cite{rt2,cari}.
\begin{eqnarray}
{\omega}_{2,sc}(q,\bmr,s)=
iq\int_0^s ds'\bigg \lbrack \sqrt{1-\frac{2m}{q^2}V(\bmr-s'\hat \bmq)}
-\sqrt{1-\frac{2m}{q^2}V(\bmr-s\hat\bmq)}\bigg \rbrack
\label{a11}
\end{eqnarray}
For $(2m/q^2V)\ll 1$, $\omega_{2,sc}$  coincides  with
$\omega_{2,V}$, Eq. (\ref{a9}). However, in classically
forbidden regions $(2m/q^2)V>1$, $\omega_{2,sc}$ describes damping,  as
the dominant imaginary part of $V_{eff}(q)$ in (\ref{a10}) is expected
to do. This will be borne out by calculations.

Whereas $\omega_{2,V}$ is strictly
proportional to $1/q$, this is no more the case for $\omega_{2,t}$
after the replacement $V\to V_{eff}(q)$.
The above manifestly introduces $q$-dependence in coefficients of the GRS
series (4a), (5b) and in particular in the BC approximation.
Taking the latter  in the cumulant form ({\ref{a7}) adds to the
blurring of the original $1/q$ dependence. This
raises the question how the response approaches its asymptotic limit.
 
We start with a theoretical analysis of the first cause of additional
$q$-dependence and focus on
$^4$He-$^4$He scattering for high lab  momenta $q$. The latter
is of distinct diffractive nature,
typical for interactions with a strong, short-range repulsion. For
those, the dominant  imaginary part  of the on-shell
scattering  amplitude $t(q)\approx {\rm Im}f(q) \propto
iq\sigma_q^{tot}$, where the total $^4$He-$^4$He cross section
$\sigma_q$ varies much  slower  than $q$  itself  \cite{rt}.
 
Without entering into details, we state that the off-shell
$\tilde t=V_{eff}$ in $\omega_2$, Eq. (\ref{a7}), can approximately be
related to the on-shell  scattering amplitude for elastic scattering.
({See ref. \cite{rte} for  a  more
extensive treatment of the parallel discussion for atomic nuclei). It can
then be shown that the rigorous proportionality of
the dominant BC FSI phase $\tilde\Omega_{2,V}\propto 1/q$ for
a smooth, bare $V$ still holds approximately for
$\tilde\Omega_{2,t}$.
 
Additional $q$-dependence is due to the use of the
cumulant representation  (\ref{a7}) but it will be small to the
extent that FSI are. In conclusion, the reduced response
described by (5) and (7) is expected to approximately preserve
the $1/q$ signature of
the dominant binary collision contribution. We shall return below to
a numerical confirmation.
 
\section{Dynamical calculations of selected MARI $^4$He data}
 
We first mention and discuss the input elements
which suffice for the BC approximation in any of the
forms described in Section II.
 
a) The atom-atom interaction $V_{Aziz}$ \cite{aziz}.
 
b) The single-atom momentum distribution $n(p,T)$
\begin{eqnarray}
n(p;T)&=&(2\pi)^3\delta(\bmp) n_0(T)+(1-n_0(T)) n^{NO}(p;T)
\nonumber\\
\frac{\rho_1(s,0;T)}{\rho}&=&n_0(T)+(1-n_0(T))\frac
{\rho_1^{NO}(s,0;T)}{\rho}
\label{a12}
\end{eqnarray}
$n_0(T\le T_c)$  is the fraction of  atoms in the condensed  state
\cite{foot1}, $n^{NO}(p;T)$ and $\rho_1(s,0;T)/\rho$ above are
respectively, the momentum distribution of the normal (uncondensed)
atoms and its Fourier transform. Path integral Monte Carlo
(PIMC) calculations have shown moderate $T$-dependence
of $n^{NO}(p;T)$ for  $T \le 4$K \cite{cepo,cep1}.
 
c) The least accessible ground state property required in the BC
approximation
is the semi-diagonal, two-body density matrix which weights the
dominant BC FSI terms in (\ref{a4c}) or (\ref{a5b}).
Calculations based on a variationally
determined groundstate wave function in the Hyper-Netted Chain (HNC)
formalism produce for $\zeta_2$, Eq. (\ref{a8}) \cite{rist,fer}
\begin{eqnarray}
\zeta_2^{HCN}(\bmr,s;\xi)&=&
g_{wd}(r)g_{wd}(|\bmr-s\hat\bmq|){\rm exp}\left[A(\bmr,s)\right]
\nonumber\\
&\approx& g_{wd}(r)g_{wd}(|\bmr-s\hat\bmq|) {\rm exp}\left [\xi
A_4(\bmr,s)\right] \nonumber\\
A_4(\bmr,s)&=&\rho \int d\bmr'
\bigg\lbrack g_{wd}(|\bmr'-s\hat\bmq|)-1 \bigg\rbrack
\bigg\lbrack g_{wd}(r')-1 \bigg\rbrack
\bigg\lbrack g(|\bmr'-\bmr|)-1 \bigg\rbrack
\label{a13}
\end{eqnarray}
$g_{wd}(r)$ is an auxiliary function  related to what in
HNC formalism is called a form factor \cite{rist}.
The function $A(\bmr,s)$ formally adds all so-called
Abe diagrams and is approximated in (13) by
the 4-body Abe diagram
$A_4(\bmr,s)$, using in addition a scaling parameter $\xi$
\cite{fer,man}.

Far less sophisticated and simpler is the GRS approximation \cite{grs1}
\begin{eqnarray}
\zeta_2^{GRS}(\bmr,s)=\sqrt{g(\bmr)g(\bmr-s\hat \bmq)},
\label{a14}
\end{eqnarray}
which interpolates $\zeta_2$ between $s=0$ and the Hartree limit
for large $s$ \cite{grs1}.
 
Both options have drawbacks and fail for instance the extended
unitarity test
\begin{eqnarray}
\int d\bmr \rho_2(\bmr-s\hat\bmq,0;\bmr,0)=(A-1)\rho_1(s,0) \,
\end{eqnarray}
\noindent
which can be written as
\begin{eqnarray}
\Xi(s)=\rho\int d\bmr\bigg(1-\zeta_2(\bmr;s)\bigg )=1
\label{a15}
\end{eqnarray}
 
Using a typical pair distribution function $g(r)$,
$\Xi(s)$ above computed with $\zeta_2^{GRS}$, Eq. (\ref{a14}),
produces values up to 1.7  for $s$=2.0 instead of the exact value 1.0,
independent of $s$ \cite{asr}.
In the HNC case, approximations involved in the evaluation of the
Abe terms (\ref{a13}) are responsible for similar deviations of
$\Xi(s)$ from 1.  The violation of  condition
(\ref{a15}) is intrinsic
in the GRS approximation (\ref{a14}), no matter what $g(r)$ is used.
 
Another important constraint is the fact that the diagonal
two-body density matrix should coincide with the pair
distribution
function: $\zeta_2(\bmr,0)=g(r)$ (\ref{a8b}).
While the GRS approximation
fulfils that condition by construction, a full evaluation of the Abe
terms is necessary in the HNC formalism.
Demanding the boundary value condition to be fulfilled in the mean,
one determines the so-called scaling
$\xi$ by  minimizing the following quantity
\begin{eqnarray}
\sigma(\xi)&=&\int d\bmr|\zeta^{HNC}(\bmr,0;\xi)-g(r)|^2
\nonumber
\end{eqnarray}
A particular
choice  of  $\zeta_2$
presumably  matters  for  medium $q$,  but  for
increasing $q\gtrsim 20\Ain$  FSI contributions
decrease  in importance relative to  the asymptotic response.
A  few-percent  spread, due to uncertainty in the choice of
$\zeta_2$, in  anyhow  small  FSI  terms will go
unnoticed.   We  thus  opted  for the  expression  (\ref{a14})  which  is
numerically much easier to handle than (\ref{a13}).
 
d) Finally, for a comparison of actual data with predictions  the latter
have to be folded  into the experimental resolution  function (ER)
$E(q,y;T)$ of  the  instrument. The $E(q,y;T)$
corresponding to the $q\ge 20 \Ain$ MARI data are given in  Azuah's
thesis \cite{azuah}
and have been fitted to the sum of two off-center gaussians. No  ER,
pertinent to lower $q$'s were available to us, thus
precluding  an analysis for $q\le 20 \Ain$.
 
Until this point we did not specify the $T$-dependence of the theoretical
responses. In fact
one  ought to
employ quantities computed for given $T$.
In fact there
exist experimental data\cite{sven} and also PIMC studies\cite{cepo,cep1}
on the $T$ dependence of the pair-distribution function $g(r,T)$.
However, in view of the above arguments we shall use the one
for $T = 0$.
By  the same  token  $\zeta_2$,  Eq. (\ref{a14}),  and
consequently FSI  effects will  be independent of  $T$.  This  leaves the
single-particle density matrices, or equivalently the momentum
distributions as the only $T-$ dependent quantities in the present
analysis.
We took $n_0(T=1.6 {\rm K})=0.087$ and  $\rho_1^{NO}(0,s;T=1.6{\rm
K})=\rho_1(0,s;T=2.3{\rm
K})$ from calculations for $T=1.54$  and 2.5K \cite{cepo,cep1}.
 
The expression for the predicted response is  therefore
\begin{eqnarray}
\phi(q,y;T\ge T_c)&=&\int \ppp n(p;T)R(q,y-p_z)
\nonumber\\
\phi(q,y;T\le T_c)&=& n_0(T)R(q,y)+[1-n_0(T)]
\int \ppp n^{NO}(p;T)R(q,y-p_z),
\label{a16}
\end{eqnarray}
which in order to enable a comparison  with data, has to be folded into
ER
\begin{eqnarray}
\phi_E(q,y;T)=\int_{-\infty}^{\infty}\,dy'\,E(q,y-y';T)\phi(q,y';T)
=(2\pi)^{-1}\int_{-\infty}^{\infty}
 ds\,e^{iys}\tilde E(q,s,T)\tilde\phi(q,s;T)
\label{a17}
\end{eqnarray}
For future reference we emphasize here that the FSI factor
$R$
is from (\ref{a7}) and (\ref{a8}) seen to be independent
of the single particle density matrix $\rho_1(s,0)/\rho$.
In particular for all but pure  hard-core interactions
\begin{eqnarray}
{\lim_{q\to\infty}} \tilde R(q,s)=1.
\label{a90}
\end{eqnarray}
 
We thus computed
the reduced response $\phi_E(q,y;T)$, for the $q= 21, 23, 25, 29
\,\Ain$ sample out of the MARI data. In view of the steady decrease of
FSI, this $q$-range and steps seems to be sufficient for our study.
We emphasize in particular the case $q=23 \Ain$,  considered
because it is the largest $q$ in the older
Argonne  data sets \cite{sos} and for it  we shall compare our
results with others.
 
We  start with  a  comparison of  our predictions  for  $T=2.3$K and  the
corresponding  data  \cite{azuah}  (Figs. 1 a-d).  The overall agreement
is very good. One notices that, whereas the central value for the
theoretical response
hardly changes for $21 \le q (\Ain)\le 29$, the data for the same,
folded in the ER, $\phi_E(q,0)$ show $q$-dependence
present in $E(q,y)$.
 
The agreement for $T=1.6$K (see Figs. 2 a-d) is slightly worse. The
slight staggering
in the central region for $q=21 \Ain$ is probably of instrumental
origin, but contrary to the $T=2.3$K case, differences in $E(q,y)$ for
$q=21, 29 \Ain$ do not explain the small
discrepancies in their central regions. We recall that exactly
the same input is used as for $T=2.3$K  and that  the only extra
parameter is the   condensate fraction $n_0(T=1.6$K).
 
We now reach our second topic.
In spite  of the fact  that no  data exist for  $q\geq 29 \Ain$,  we have
extended calculations up  to $q= 300 \Ain$.
The  purpose of  the  exercise  is to  obtain
$theoretical$  information  on  the  approach  of  the  response  to  its
asymptotic limit.

In  Fig. 3a we present  the part $\phi^{even}(q,y,T=2.3{\rm  K})$ of  the
predicted response,  computed with (\ref{a7}), (\ref{a10})  which is even
in $y$. Even in the  wings out to
$  y \approx  3.5 \Ain$  those coincide
within  $1\%$  amongst  themselves   and  with  the  asymptotic  response
$F_0(y)$,  Eq.  (\ref{a4b}).   The  above appears  hardly  changed,  when
predictions for $20\le q(\Ain)\le 29$ are included: only in the immediate
neighborhood of $y=0$, is there a $\lesssim 2.5 \%$ difference.
 
In Fig.  3b we  show $q\phi^{odd}(q,y,T=2.3{\rm  K})$ which  contains the
part of the response  which is odd in $y$. Having included the
$q$ signature  of the dominant FS some residual
$q$ dependence is apparent
in the  extrema as  well as in  the wings, but  the conclusion  is clear:
Neither  the strong  short-range repulsion  in the  atom-atom interaction
which  forces the use of $V_{eff}(q)=\tilde t(q)$, nor the effect
of the cumulant
representation, much changes the $1/q$ signature of the dominant FSI term
in the GRS series (\ref{a4a}) for  smooth $V$.  The above agrees with our
arguments in Section II and with our previous
results \cite{rt}.
 
All reported predictions are based on  the use of the $t$-matrix, i.e. on
(\ref{a10}). In Section  II we  also mentioned
a  semi-classical approximation (\ref{a11}) for FSI
and found that, except for small $s$, there are
considerable differences  between the  BC   phases, calculated
by means of (\ref{a11})   and (\ref{a10}).
 
Ultimately excellent   agreement  is   obtained   between   the
corresponding  responses,   computed  with  (\ref{a5})   and  (\ref{a7}).
Clearly  both the  $t$-matrix  and the  semi-classical method  accurately
describe the binary collision phase in the salient region just inside the
classically  forbidden  region.   Contributions from  deeper  penetration
distances are strongly suppressed.
 
We conclude this Section, by a
discussion and comparison of  predictions  for $q=23\Ain$
by other authors.
Since the various studies  refer to different
$T$  and data  have  been  taken at  different  instruments, the  natural
quantity  to   compare  is  the   FSI  factor  $R(q,y)$  assumed   to  be
$T$-independent.
 
We start with predictions by Mazzanti et al \cite{fer} which are
based on exactly  the same BC approximation in  the $V_{eff}$
version (\ref{a10}), employing  however
the variationally derived $\zeta_2^{HNC}$, Eq. (\ref{a13}).
Next we mention Silver \cite{sil} who  used, what amounts to the cumulant
form (\ref{a7}) with $\omega_2\to \omega_{2,t}$ and $\zeta_2
\to g_2$, the pair distribution function.  In his  Hard Core Perturbation
Theory  he disregarded
the second part of the total phase $\chi$, Eq. (\ref{a4d}),
which is only permissible for a $pure$ hard-core interaction.
However, Silver actually constructed the off-shell $t$-matrix in
(\ref{a10}), corresponding to the first part  in  (\ref{a4d}) from JWKB
partial wave phase shifts for a realistic $V$, which in addition to
strong short-range repulsion also included
attractive components. Nevertheless, he neglected the second component
in (4) which does not vanish when an attraction is present.
We conclude with a path-integral  method by Carraro and Koonin, who
computed high-$q$ FSI  using a fixed scattering  approximation for the
entire  system with  a large,  finite number  of atoms  \cite{cako}.  The
method requires the parallel calculation of the ground
state wave function
in order  to construct the  $N$-body density matrix, diagonal  except for
one particle, with $N$ the number of atoms in the sample and which
averages the response for fixed scatterers.
 
Results for $R(q,y)$ for all cases discussed are assembled in
Fig.  4  and  show  occasionally substantial  differences.
However,  those  are  considerably  smoothened  by  the  single  particle
momentum distribution  (Eq. (\ref{a5})) or density  matrix in (\ref{a6}),
and ultimately produce quite  similar responses
\cite{sos,fer}.

\section{Cumulant expansions of the response.}
 
We consider below a method which has extensively been applied
in  the  past  \cite{sears}  before  the rediscovery  of  the  $1/q$  GRS
expansion of the  response (\ref{a4a}) \cite{grs1}.
Recently it has been brought to
the fore again in an attempt to  parameterize data without  the
intervention of  a theory.   The method uses  cumulant expansions  of the
Fourier  transforms of  the  separate  asymptotic and  FSI  parts of  the
response (\ref{a5b}),  with coefficient  functions related to  moments of
the response \cite{glyde}
\begin{mathletters}
\label {a18}
\begin{eqnarray}
{\tilde \phi}(q,s)&=&{\rm exp}\bigg\lbrack \sum_{n \ge 2}\frac
{(-is)^n}{n!}{\bar \mu}_n(q)\bigg \rbrack
\label{a18a}\\
\tilde F_0(s)&=&{\rm exp}\bigg \lbrack \sum_{n \ge 2}\frac
{(-is)^n}{n!}{\bar \alpha}_n\bigg \rbrack
\label{a18b}\\
{\tilde R}(q,s)={\rm exp}[\tilde \Omega(q,s)]
&=&{\rm exp}\bigg \lbrack \sum_{n \ge 3}\frac
{(-is)^n}{n!}{\bar \beta}_n(q)\bigg \rbrack
\label{a18c}
\end{eqnarray}
\end{mathletters}
Using   (\ref{a5b}) the various coefficient functions are
related by ${\bar \mu}_n(q)={\bar \alpha}_n+{\bar  \beta}_n(q)$
\cite{foot2}.
Data for the response $\phi_E^{exp}$ are then compared
with the parameterization (\ref{a18})
\begin{eqnarray}
\phi_E^{exp}(q,y)\Longleftrightarrow \frac{1}{\pi}{\rm Re}\int^{\infty}_0
ds e^{iys}\tilde\phi_E(s,[\bar\mu_n(q)]),
\label{a19}
\end{eqnarray}
where, as in  (\ref{a17})  the  right-hand side  in  (\ref{a18a})
includes ER.   In principle, no a priori
knowledge, or even  meaning  of the  cumulant coefficient
functions $\mu_n(q)$ is necessary for a search.  However, it is
natural to use motivated initial  values such as calculated ones.
 
Consider the normal fluid, in which case the
$q$-independent cumulant coefficients ${\bar \alpha}_n$ originating
in  the asymptotic part of $\tilde\phi(q,s)$ can simply be expressed in
terms   of averages of even powers of the momentum of an atom, e.g.
\begin{eqnarray}
\bar\alpha_2&=&\frac{1}{3} \langle p^2 \rangle
\nonumber\\
\bar\alpha_4&=&\frac{1}{5} \langle p^4\rangle-
\frac{1}{3}\langle p^2\rangle^2 \,\,\,,{\rm  etc.}
\label{a20}
\end{eqnarray}
The $q$-dependent coefficients $\bar\beta_n(q)$  relate to
the FSI factors  $\tilde R(q,s),\tilde\Omega(q,s)$ in (\ref{a5b})
and may be written as
\cite{glyde}
\begin{eqnarray}
{\bar \beta}_n(q)=\sum_{m=1}^{[(n-1)/2]}
\bigg(\frac{1}{v_q}\bigg )^{n-2m} {\bar \beta}_{n,2m}
=\sum_{m=1}^{[(n-1)/2]} \bigg (\frac{1}{q^*}\bigg )^{n-2m} z_{n,2m}
\label{a21}
\end{eqnarray}
For convenience we use
$q^*$, the momentum transfer parameter $q$ in units of 10$\Ain$.
Eq. (\ref{a21}) displays $q-$dependence and defines coefficients
$z_{n,2m}$, which have the same dimensions and
may be expressed in the  same units as $\bar \beta_n$.
The above are
operators for dynamical variables of the system, averaged
over $diagonal \,\,l$-body density matrices and  their derivatives
of order $l\leq n$.  For the lowest order cumulant
coefficient functions one has
\begin{mathletters}
\label{a22}
\begin{eqnarray}
\bar\beta_3(q)&=&6\,{\lim_{s\to 0}}\,[{\rm Im}\,\tilde\Omega_2(q,s)/s^3]
\label{a22a}\\
&=& \frac{1}{6v_q} \langle\nabla^2 V \rangle_{\rho_2}
\label{a22b}\\
\bar\beta_4(q)&=&24\,{\lim_{s\to 0}}\,[{\rm Re}\,\tilde\Omega_2(q,s)/s^4]
\label{a22c}\\
\bar\beta_5(q)&=&\frac{1}{v_q}\bar\beta _{54}
+\bigg (\frac{1}{v_q}\bigg) ^3
\bar \beta_{52}
\label{a22d}
\end{eqnarray}
\end{mathletters}
Contrary to the GRS series in $1/q$, with coefficients depending on
nearly-diagonal $n$-particle density matrices
$\rho_n(\bmr_1-s\hat\bmq,\bmr_2,...\bmr_n;\bmr_1,...\bmr_n)$,
the moment approach underlying
the cumulant expansion does not produce a systematic $q$-dependence
of the coefficient functions. For instance, $all$
$\bar\beta_n(q)$ with odd $n$ contain FSI
contributions $\propto 1/q$. Eqs. (\ref{a22a}) and (\ref{a22d})
illustrate this for the dominant FSI  coefficient functions
$\bar\beta_n(q)$. Thus $\bar\beta_3(q)\propto 1/q$
draws exclusively on the BC  contribution (\ref{a7}), while the two
components of $\bar \beta_5$ are proportional to $1/q$
and $(1/q)^3$ due to binary, respectively
higher order collision contributions, etc. It is the expansion of the
semi-diagonal $\rho_2(\bmr_1-s\hat\bmq,\bmr_j;\bmr_1,\bmr_j)$
in $s$ which produces an infinite number of contributions
$\bar\beta_{2n+1}(q)$, all of which have parts
$\propto 1/q$ with coefficients depending on
$diagonal$ densities (cf. (\ref{a22b})) and their derivatives.
 
We start with the threshold behavior of the computed
BC FSI phase $\tilde\Omega_2(q,s)$, Eqs.
(\ref{a7}), (\ref{a10}) and in particular of its imaginary part
which, according to (\ref{a22a}), produces the  dominant FSI parameter
$\bar\beta_3(q)$. We checked that, within a few percent
$q\,{\rm  Im} \tilde\Omega_2(q,s)$ is out to $s\approx 0.8\AAr$
independent of $q$. In
particular we could extract the theoretical threshold
value $z_3=q^*\bar\beta_3^{\Omega_2}(q)=
6q^*\,\,{\lim_{s\to 0}[{\rm  Im}\tilde\Omega_2(q,s)/s^3]}=
0.555 \AAr^{-3}$ which over the entire  range $q\le  300\Ain$ is
to better than  $ 1\%  $ independent of $q$. For the above reason one
cannot determine the next order odd-$n$ coefficients $z_5$ with
reasonable precision.
 
Regarding parameters of even order in $n$ one finds
from Re $\tilde\Omega_2(q,s)$, Eq. (\ref{a22c}) for the leading
coefficient $z_{42}=q^{*2}\bar\beta_4(q)=-2.26 \AAr^{-4}$.
It appears that for even $n$, the next-to-leading
$z_{64} \approx q^{*2}\bar\beta_{64}(q)
\approx -31\AAr^{-6}$ is reasonably well defined.
 
We return to Eq. (\ref{a22b}) which seems to provide an
independent
way to calculate $\bar\beta_3(q)$.
However, one can  actually derive it from  (\ref{a22a}), using (\ref{a7})
for $\tilde\Omega_2$ in either version  (\ref{a9}) or (\ref{a10}) for the
generalized eikonal  phase $\tilde\omega(q,s)$.   It holds  for arbitrary
semi-diagonal $\rho_2$ which  exactly satisfies (\ref{a8b}) \cite{foot3}.

Using the same $g(r)$ as in $\zeta_2^{GRS}$ (\ref{a14})
we compute
$q^*\bar\beta_3^{{\nabla^2}V}(q)=0.56\AAr^{-3}$.
The agreement between $q^*\bar\beta_3^{{\nabla^2}V}$ and
$q^*\bar\beta_3^{\tilde\Omega}$ is very good, especially
in view of
the sensitivity of $q^*\bar\beta_3^{{\nabla^2}V}$ on the
precise shape of $g(r)$ where the Laplacian of $V$is large.
The extraordinary stability of the
extracted $q^*\bar\beta_3^{\tilde\Omega}$ confirms the numerical
consistency of the calculation.
 
As has been
mentioned before, all FSI functions have been assumed to be
$T$-independent and we have used values for $T=0$. In order to estimate
the influence of the temperature we have also calculated
$q^*\bar\beta_3^{{\nabla^2}V}$ using a $g(r)$ obtained with PIMC at
$T=2.8 K$  \cite{cepo}. The result
$q^*\bar\beta_3^{{\nabla^2}V}(T=2.8 K)=0.47 \AA^{-3}$ confirms its
sensitivity on the precise shape of $g(r)$.
 
We now report several results for cumulant fits (\ref{a19}), obtained
with the CERN MINUIT  code and applied to the ten  $T=2.3$K data sets in
the  range  $21\le  q(\Ain)\le  29$ from Azuah's  thesis  \cite{azuah}.
We first  note that the integrated
strengths  of the data there
appear for all $q$ to be  approximately 1.4$\%$ in excess of the
exact  result  1. By construction that
demand is exactly fulfilled by a cumulant expansion (\ref{a18a}) in, no
matter what  approximation.  We also  considered data cut  at $ y \approx
3.0-3.4 \Ain$ where statistical noise  in anyhow very small responses may
cause those to have negative values.   The latter appear to hardly affect
the extracted parameters.
 
Next, the above source of information does not contain numerical data and
ER for the lower  $q$ data .  As a consequence we  had to limit ourselves
to  a small  data base  which  is bound  to influence  the FSI  parameter
functions which increases with decreasing $q$.
 
Our results for $T$=2.3K are assembled in Table I. We entered in column 2
theoretical  values for the parameters, calculated   as  indicated
,  or set to  0 when impossible to evaluate
reliably.
Notice that the negative$z_6$, obtained from a limited $s-$range,
will generate a non bound $\tilde R(q,s)$, barring
a convergent Fourier transform
of $\tilde R(q,s)$. Actually we
did not
consider $z_6$ in the fits and restrict ourselves to a maximum of seven
parameters as is discussed below.
 
Column 3  is the result  of a  7-parameter fit for  cumulant coefficients
functions with  theoretical $q$-dependence and  encompasses therefore
all ten data sets in our $q$-range.  Columns 4 and 5 are 6-, respectively
5-parameter fits like  in column 3, when  first $q^*\bar\beta_3(q)$ alone
and  then  also  $q^{*2}\bar\beta_4(q)$  are held  at  their  theoretical
values.
 
We now  reach the extraction  of $\bar\mu_n(q)$  for each $q$  and the
$determination$, as opposed to the above $assumption$, of their
$q$-dependence.
This appeared  to be impossible for the limited data set available to us.
However, Glyde et al had  access to far more data and
we enter their results in the last column taken from Table I of Ref.
\cite{glyde1}. As the reported $z_4$ is zero, a positive value of
$z_6$
is required to give a convergent Fourier transform of $\tilde R(q,s)$.
 
We hinted in the Introduction,  that the
misfit of FSI parameters may reflect on the meaning
of the parameters from which one builds the momentum distribution $n(p)$.
In fact, the above was the main reason why we  repeated the fit of the
cumulant expansion coefficients.  Table I shows even smaller values for
$z_3=q^{*}\beta_3(q)$
than reported
in \cite{glyde,glyde1}. Although the fitted $z_3$  appears  strongly
correlated to other parameters, we remained worried by the misfit.
The above observation led us to ponder about the very
possibility to obtain reliable values for the cumulant coefficients for
the response.
 
In fact, we  found the results of column  5 in Table
I most telling.  Fixing the  dominant FSI parameters $z_{3,4}$ at their
theoretical  values, one  expects  $\bar\alpha_4,\bar\alpha_6$ to  settle
close to their starting values.  This appears not to be the case.  In
fact,  the above  happens  to  produce the  worst  agreement between  the
reconstructed and computed $n(p)$ to be presented below.
 
To understand  the  above, we  turn  again  to the  calculated
$\tilde\Omega(q,s)$ which we recall, produces very good fits to the data.
Although  not  called for  in  calculations  using  a dynamic  theory,  a
comparison of (\ref{a5b})  and   (\ref{a18c})  shows  that the
expansion  of $\tilde\Omega(q,s)$   in  $s$  produces in   principle
the   cumulant
coefficients and in particular  the FSI parameters $\bar\beta_n(q)
$. Examples are (\ref{a22a}) and (\ref{a22c}).
 
The  complete  cumulant series  is  of  course  equivalent to  the  exact
$\tilde\Omega$,  but  a  truncation  at  some  $n$  obviously  reproduces
behavior up to some relatively low  $s$.  The crucial question is to what
order one  should go and the  answer clearly depend on  the effective $s$
range  of  each  of  the  component  factors.   Those  are  according  to
(\ref{a5b}),   (\ref{a17})  and   (\ref{a19})  the   single-atom  density
distribution  $\rho_1(0,s)/\rho$,  the  FSI  interaction  factor  $\tilde
R(q,s)$  (or  $\tilde \Omega(q,s)$)  and  the  Fourier transform  $\tilde
E(q,s)$ of the ER function.
 
In the tests  to be reported we compare not  data, but theoretical values
of  the input  factors $\rho_1(0,s),  \tilde R,\tilde  \Omega$ and  their
cumulant approximations  $\tilde R^{cu,n},\tilde\Omega^{cu,n}$ to
order  $n$.  Fig.  5a  shows  over the  relevant  $s$-range
reasonable adequacy  of the cumulant expansion  for $\rho_1(0,s)/\rho$ to
order   $n=6$.    Next,   Figs.   5b,c   give  Re   and   Im   parts   of
$\tilde\Omega(q,s)$,  respectively $\tilde  R(q,s)$ $q=23\Ain$  and shows
that the  FSI cumulant  series, truncated at  $n=5$, rapidly  falls short
from of the computed functions for $s\ge 1\Ain$.
Inclusion of
the above mentioned, well-determined
$z_6\approx z_{64}$ affects only Re$\tilde\Omega$, (cf.
Eq. (\ref{a18b}). It
extends the agreement between the  calculated and cumulant expansions for
the  two FSI  functions over a modest additional
$s$-range, but does not prevent rapid deterioration of $\tilde R(q,s),
\tilde \Omega(q,s)$ from 1.2 $\Ain$ on. As mentioned before the estimated
$z_6$ gives rise to a non-bound $\tilde R(q,s)$.  Since terms in the
cumulant expansions (\ref{a22}) have alternating signs
of powers in $s$, it  requires knowledge of higher order $\beta_n$ in
order to establish the proper order of the expansion for
given $s$ range.
 
We return to the interpretation of our results.
The cited parameters for finite$-n$ expansions indeed reproduce very good
fits to data, but have no resemblance to the values of the
parameters from  actual (in case theoretical)  FSI factors
$\tilde R(q,s),\tilde\Omega(q,s)$.  The failure is due to the
insufficiency of low order
cumulant representations $\tilde R^{cu,n},\tilde\Omega^{cu,n}$
and will only
disappear if for sufficiently high $n$, FSI functions will coincide
with  their cumulant expansion out to the relevant medium $s$.
 
The  above is  to  our  opinion the  source  of  the discrepancy  between
calculated  and  extracted  FSI  parameters.  In  spite  of  correlations
between fitted parameters,  there is sufficient meaning  to the extracted
FSI to  justify the above conclusion.   Were it not for  the overwhelming
role  of  the asymptotic  part  and  its  representation by  a  truncated
cumulant series, one could not trust the extracted $\bar\alpha_n$.
 
The sensitivity of the fitting procedure can be judged from Fig. 6, where
we  reconstruct  single-atom  momentum  distributions  from  the  various
$\alpha_n$.  Dense dots,  short dashes and spaced dots  are results using
columns 3-5 from Table I.  Long dashes are from the last column, i.e. the
fit of  Glyde et  al.  All  fits are roughly  of comparable  quality when
judged  as  to their  capacity  to  reproduce the  computed  distribution
$n(p,T=2.5$K)  \cite{cepo},  marked  by  circles.   We  recall  that  the
theoretical $n(p)$ has been used to  compute $\alpha_n$ in column 2 which
produces the drawn curve  in Fig. 6 .  The moderate  misfit for large $s$
of  the  finite  order  cumulant  expansion  for  $\rho_1(0,s)$  explains
deviations from the underlying $n(p)$.

\section{Summary and conclusions.}
 
We addressed above three topics  regarding the response of liquid $^4$He,
retrieved from inclusive scattering of neutrons.  Using dynamics we first
made predictions for the recent  MARI data, taken at
 temperatures both
below  and above  $T_c$.  We  analysed  the range  of momentum  transfers
$21\le q(\Ain)\le 29$,  for which we had available  data and experimental
resolution and obtained good agreement  with experiment.
 
A second topic  was the approach of the response  to its asymptotic limit
in $q$ for  fixed scaling variable $y$.  For  smooth interactions between
constituents, that approach is rigorously  $\propto 1/q$, but the same is
not  guaranteed when  a strong  short-range repulsion  is present  in the
atom-atom interaction  $V$.  We  investigated theoretically  the response
for $q\le 300\Ain$ for the actual  $V$ with its short range repulsion and
found re-confirmed our findings from a  few years ago based on medium-$q$
data:  Final   State  Interaction  contributions,  over   and  above  the
asymptotic part still decrease approximately as $1/q$.
 
Regarding  the $q$-values  for which  the response  has been  measured we
repeat what was already evident from  the older data \cite{sos}: there is
no additional information to be retrieved  by increasing $q$ by less than
20-30\% and no new information at all at very high $q$.
 
Our last topic was a re-fit of the expansion coefficient functions, which
parametrize  data.  A  previous analysis  led to  a single  atom momentum
distribution  in  good agreement  with  computed  $n(p,T)$, but  did  not
produce  the  main  FSI  coefficient  function.   Its  influence  in  the
large-$q$ region  is too  marginal, to  be extracted  from the  data with
present accuracies. This does not change our judgement that little can be
added to $understanding$ the data on the response of liquid $^4$He
at high momentum.

In spite of minor discrepancies, a  summary of the treatment of the first
two topics is definitely positive.   Presumably for no atomic, molecular,
nuclear or sub-nuclear  system for which the response  has been measured,
has  one reached  asymptotia as  clearly and  as well  understood as  for
liquid $^4$He.  Of course $'$asymptotia$'$ is not simply the mathematical
limit $q\to  \infty$ for fixed  $y$.  Increasing $q$  requires increasing
beam energy, ultimately beyond the ionization energy $\approx $ 39 eV.  A
description of  the response  then requires  the inclusion  of additional
electronic degrees of freedom to translational ones.
 
Finally we venture an outlook for future explorations of inclusive
scattering which almost certainly implies extension of experiments to
larger momentum transfers $q$. Our judgement is based on Eqs. (\ref{a5b})
and (\ref{a7}). In that representation for the response,
the FSI factor $\tilde R(q,s)$ does not depend on the single particle
density $\rho_1(0,s)$ and in particular not on the condensate fraction
for $T\le T_c$. Together with (\ref{a90}) this implies the universality
of the asymptotic limit (\ref{a4b}) under all circumstances. One then
tends to believe that present experiments and
theory seem  to have exhausted  the search for information,  contained in
the response of liquid $^4$He at high momentum.

\section{Acknowledgments.}
ASR thanks the Department d'Estructura
e Constituents de la Materia, Universtat De Barcelona for hospitality and
support. He also takes pleasure in thanking H. Glyde for an informative
correspondence.  The authors are grateful to D. Ceperley for supplying
$g(r)$ at finite temperatures.
This work was partially supported by DGICYT (Spain) Grant No. PB95-0761.

\newpage
 
{Figure Captions}
 
Figs.   1a-d.
Predictions for the response (\ref{a5}), (\ref{a7})  of  liquid $^4$He at
$T=2.3$K for  $q=21, 23, 25, 29 \Ain$ from  (\ref{a7}) using
the  ladder  approximation  (\ref{a10})  for  binary  collisions.   Those
results have been folded in experimental resolutions from Azuah's thesis.
The same is the source of data \cite{azuah}.
 
Figs 2a-d.
Same as Figs. 1a-d, for $T=1.6$K, computed for $n_0(T=1.6$K)=0.087.
 
Fig. 3a.
$\phi^{even}(q,y)$, part  of the response even in $y$  for
$21\le q(\Ain)\le 300$. For $q\ge 25\Ain$ those can not be
distinguished from the asymptotic limit $F_0(y)$, Eq. (\ref{a6}).
 
Fig.  3b.
$q \phi^{odd}(q,y)$, $q$ times the response, odd in $y$ for
$21\le q(\Ain)\le 300$.
 
Fig. 4 .
The FSI function $R(q=23 \Ain,y)$ in (\ref{a5a}), computed
with the GRS and HNC choices for $\rho_2,\zeta_2$ as well as for other
descriptions. The drawn line is our result.
Long dashes, short dashes and dots are from references
\cite{fer,cako,sil}
 
Figs. 5.
a) Comparison of of the single atom
density matrix  $\rho_1(0,s)/\rho$ and its 6th order cumulant expansion.
b) Real and imaginary parts of the FSI phase $\tilde\Omega(q,s)$
($q=21 \Ain$) from theory and represented by a 5-th, respectively 6-th
order cumulant expansion; c) the same for  the FSI factor
$\tilde R(q,s)$.
 
Fig. 6.
The single atom momentum distribution for $T$=2.3 K, reconstructed
from the various cumulant fits, assembled in  Table I, including the one
from theoretical starting values. The curves correspond for
increasing column number of the column in Table I
including the fit by Glyde et al.
The circles are  values   calculated in \cite{cepo}.
 

\newpage
\begin{center}
{\bf Table I}
\vskip 1cm
\begin{tabular}{|c|c|c|c|c|c|}
\hline
Cumulant            & computed        & 7-parameter
& 6-parameter       & 5-parameter
&  Glyde et al.          \\
coefficient         & starting        & fit for prescribed
& fit; fixed        & fit; fixed
& \cite{glyde,glyde1}    \\
                    & values          & $q$-dependence
& $q^*\bar\beta_3$& $q^*\bar\beta_3,q^{*2}\bar\beta_4$
&                     \\
\hline
$\bar\alpha_2  (\AAr^{-2})$        & 0.916
& 0.910 & 0.913                    & 0.914        & 0.897\\
$\bar\alpha_4  (\AAr^{-4})$        & 0.470 & 0.553
& 0.594 & 0.781& 0.46\\
$\bar\alpha_6  (\AAr^{-6})$& 0.337 & 0.535 & 0.613
& 0.700 &0.38 \\
\hline
$q^* \bar\beta_3(\AAr^{-3})$ & 0.555
& 0.237 & 0.555   & 0.555  &  0.33 \\
$q^{*2}\bar\beta_4(\AAr^{-4})$ &-2.268
& -0.698  & -0.993&-2.268 &0\\
$q^*\bar\beta_{54}(\AAr^{-5})$ &      0
& 0.416   & 0.851&  0.615 & 0  \\
$q^{*3}\bar\beta_{52}(\AAr^{-6})$ & 0& -0.152
& 1.32 & 3.23  & 0.201\\
$q^{*2}\bar\beta_{64}(\AAr^{-6})$ & $\approx$ -31.0 & --- & ---
& --- &  1.539\\
\hline
\end{tabular}
\end{center}
\vskip 1cm
Cumulant coefficient functions from data  at $T$=2.3K.  The second column
gives theoretical values.  The third  column gives
7-parameter  fits with  prescribed $q^*$  behavior from  (renormalized)
data.  Column  4 give fits  when $q^*\bar\beta_3(q)$  is
fixed  at  its  starting  values.   Column 5 are  fits  if  in  addition
$q^{*2}\beta_4(q)$ is fixed.
In the last column are results by Glyde et al, who
fitted $\mu_n(q)$ from a large set of data for different $q$.

\begin{references}
 
\bibitem{and1}
K.H. Andersen et al, Physica B180/181, 865 (1992); $ibid$
Physica B197, 198 (1994).
 
\bibitem{azuah} R.T. Azuah, PhD Thesis, Univ. of Keele (1994).
 
\bibitem{sos} T.R. Sosnick, W.M. Snow, R.N. Silver and P.E. Sokol,
Phys. Rev. B41, 11185 (1990); $ibid$ B43, 216 (1990).
 
\bibitem{sil}
R.N. Silver, Phys. Rev. B37, 3794 (1988); $ibid$ B38, 2283
(1988); $ibid$  B39, 4022 (1989).
 
\bibitem{rt}
A.S. Rinat and M.F. Taragin, Phys. Rev. B41, 4247 (1990).
 
\bibitem{rt1}
A.S. Rinat, Phys. Rev. B42, 9944 (1990).
 
\bibitem{cako}
C. Carraro and S.E. Koonin, Phys. Rev.  B41, 674 (1990);  Phys. Rev.
Letters 65, 2792 (1990).
 
\bibitem{fer}
F. Mazzanti, J. Boronat and A. Polls, Phys. Rev. B53, 5661 (1996).
 
\bibitem{glyde}
H.R. Glyde, Phys. Rev. B50, 6726 (1994).
 
\bibitem{glyde1}
H.R. Glyde, R.T. Azuah, K.H. Anderson and W.G. Stirling, to be
published in Condensed Matter Theory 10, (1995);
R.T. Azuah, W.G. Stirling, H.R. Glyde, P.E. Sokol and S.M. Bennington,
Phys. Rev. B51, 605 (1995).
 
\bibitem{ken}
K.H. Andersen, W.G. Stirling and H.R. Glyde, to be published.
 
\bibitem{cepo}
D.M. Ceperley and E.L. Pollock, Can. J. Phys.  65, 1416 (1987) and
private communication.
 
\bibitem{wp}
 P.A. Whitlock and R.M. Panoff, Can. J. Phys 65, 1409 (1987).
 
\bibitem{grs1}
H.A. Gersch, L.J. Rodriguez and Phil N. Smith, Phys. Rev.
A5, 1547 (1973).
 
\bibitem{west}
G.B.  West, Phys. Reports C18, 264 (1975).
 
\bibitem{grs2}
H.A. Gersch and L.J. Rodriguez, Phys. Rev. A8, 905 (1973).
 
\bibitem{bespro}
J. Besprosvany, Phys. Rev. B43, 10070 (1991).
 
\bibitem{rt2}
A.S. Rinat and M.F. Taragin, Phys. Lett, B267, 447 (1991).
 
\bibitem{cari}
C. Carraro and A.S, Rinat, Phys. Rev. B45, 2945 (1992).
 
\bibitem{rte}
A.S. Rinat and M.F. Taragin, Nucl. Phys. A598, 349
(1996).
 
\bibitem{aziz}
R.A. Aziz, V.P.S. Nain, J.S. Carley, W.L. Taylor and  J.T. McConville, J.
Chem. Phys. 70, 4330 (1979).
 
\bibitem{foot1}
One knows more details of the shape of the momentum
distribution of the condensate fraction \cite{gav} than implicit in
(\ref{a12}), but the latter suffices for our limited purposes below.
 
\bibitem{gav}
J. Gavoret and P. Nozieres, Ann. of Phys. (N.Y.),
28, 349 (1964).
 
\bibitem{cep1}
D.M. Ceperley in $'$ Momentum Distributions $'$, Plenum Press,
Eds. R.N. Silver and P.E. Sokol, p. 71;
 
 
\bibitem{rist}
M.L. Ristig and J.W. Clark, Phys. Rev. B40,4355 (1989)
 
\bibitem{man}
Q. N. Usmani, B. Friedman, and V.R. Pandharipande,
Phys. Rev. B{\bf 25}, 4502 (1982).
 
 
\bibitem{asr}
A.S. Rinat, Phys. Rev. B40, 6625 (1990).
 
\bibitem{sven}
E.C. Svenson, V.F. Sears, A.D.B. Woods and P. Martel, Phys. Rev.
B21, 3638 (1981).
 
\bibitem{sears}
V.F. Sears, Phys. Rev. B30, 44 (1984) and references contained therein.
 
\bibitem{foot2}
The quest for manageable magnitudes
for the  coefficients  $\bar\alpha_n,\bar\beta_n(q),  \bar
\mu_n(q)$ which have dimensions  $l^{-n}$, has occasionally led to the
introduction of  artificial  combinations  of units which varied with
the order $n$ \cite{glyde,glyde1}.  A possible candidate for such a
length is  the mean free path of an atom in the medium.
From measured, weakly oscillating total atom-atom cross
sections in the relevant $q$-range \cite{felt1} one estimates
$2\lambda=[\rho\sigma_q^{tot}/2]^{-1}\approx 1.6{\rm \AAr}$, which may
serve as a scale for the recoil length $s$.
For instance, Eq.\ref{a18a}) would become $${\tilde \phi}(q,s)=
{\rm exp}\bigg\lbrack \sum_{n \ge 2}\frac
{(-is)^n}{n!}{\bar \mu}_n(q)\bigg \rbrack
={\rm exp}\bigg\lbrack \sum_{n \ge 2}
(-is/2\lambda)^n\,M_n(q)\bigg \rbrack,$$
with the, now dimensionless $M_n(q)=(2\lambda)^n\mu_n(q)/n!$.
The thus defined coefficients permit an estimate of their relative size.
 
\bibitem{felt1}
R. Feltgen, H. Pauly, F. Torello and H. Veymeyer, Phys. Rev. Lett.
30, 820 (1973); R. Feltgen, H. Kirst, K.A. Koehler, H. Pauly and
F. Torello, Journ. of Chem. Phys. 76, 2360 (1982).
 
\bibitem{foot3}
The  above  seems to  run
counter a statement by Glyde (text after (Eqs. (19.28) in \cite{glyde2}))
that evaluations starting  from (\ref{a5b}) forcibly lead to  a factor -2
difference between  (\ref{a22a}) and (\ref{a22b}).  The  latter is easily
shown to be  caused by the use  of a diagonal instead  of a semi-diagonal
$\rho_2$ in  $\tilde\Omega_2$. (The semi-classical approximation  (12) is
not  exactly  in  that  category  and for  it,  the  right-hand  side  of
(\ref{a22b}) has  to be  supplemented by a  small correction  of relative
order  ${\cal   O}(q^{-2})$.
 
\bibitem{glyde2}
H.R. Glyde, 'Excitations in liquid and solid He', p.340 (Oxford
University Press, 1994).
 
\end{references}
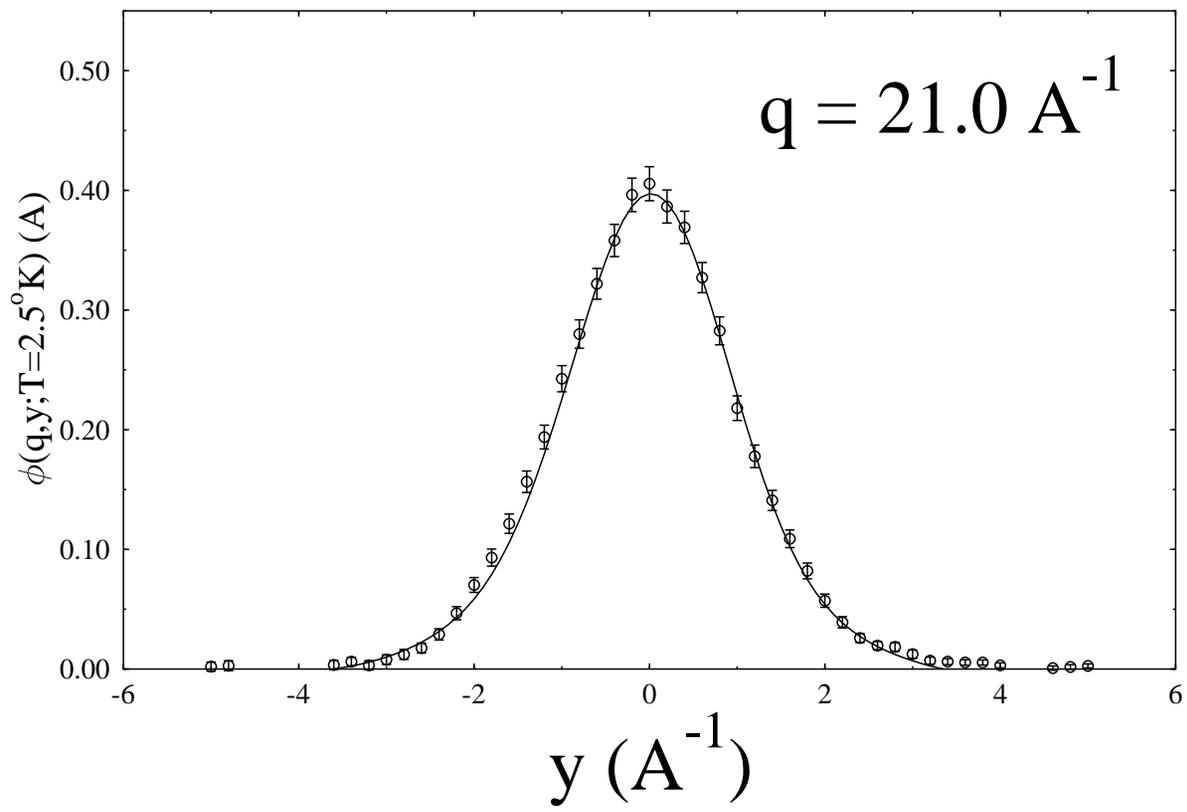
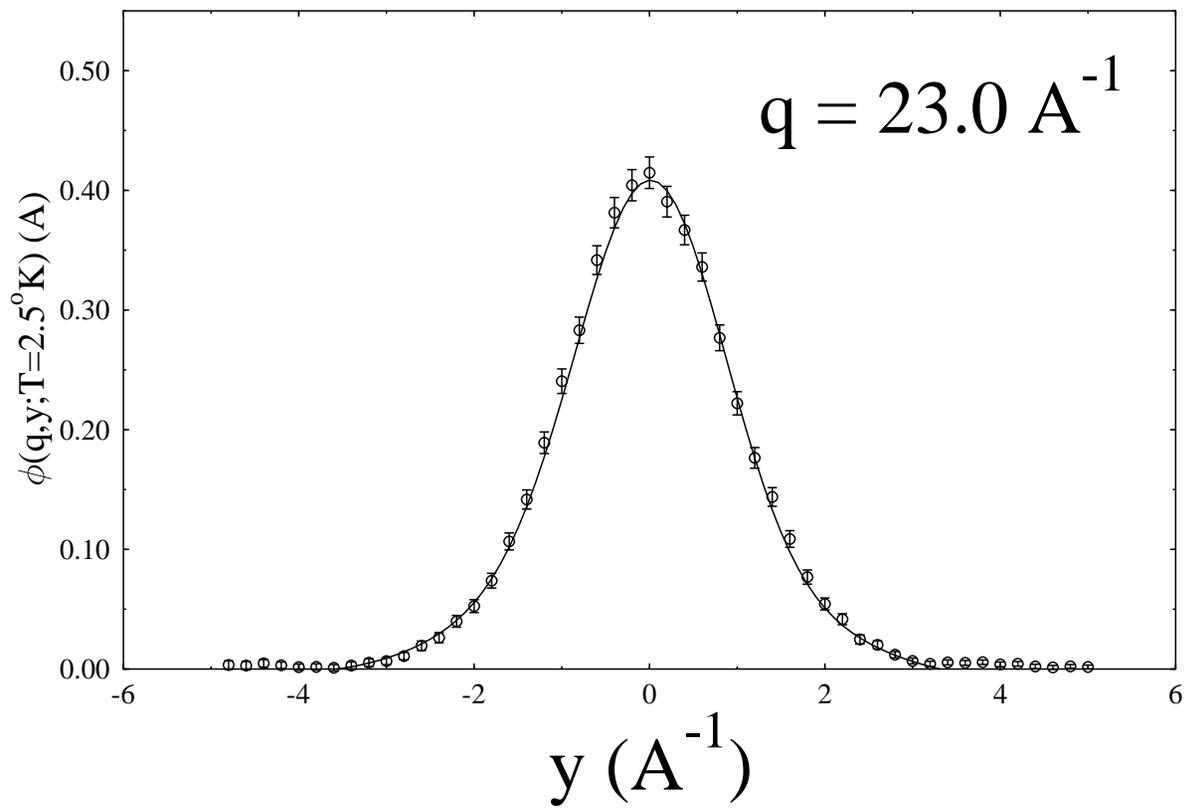
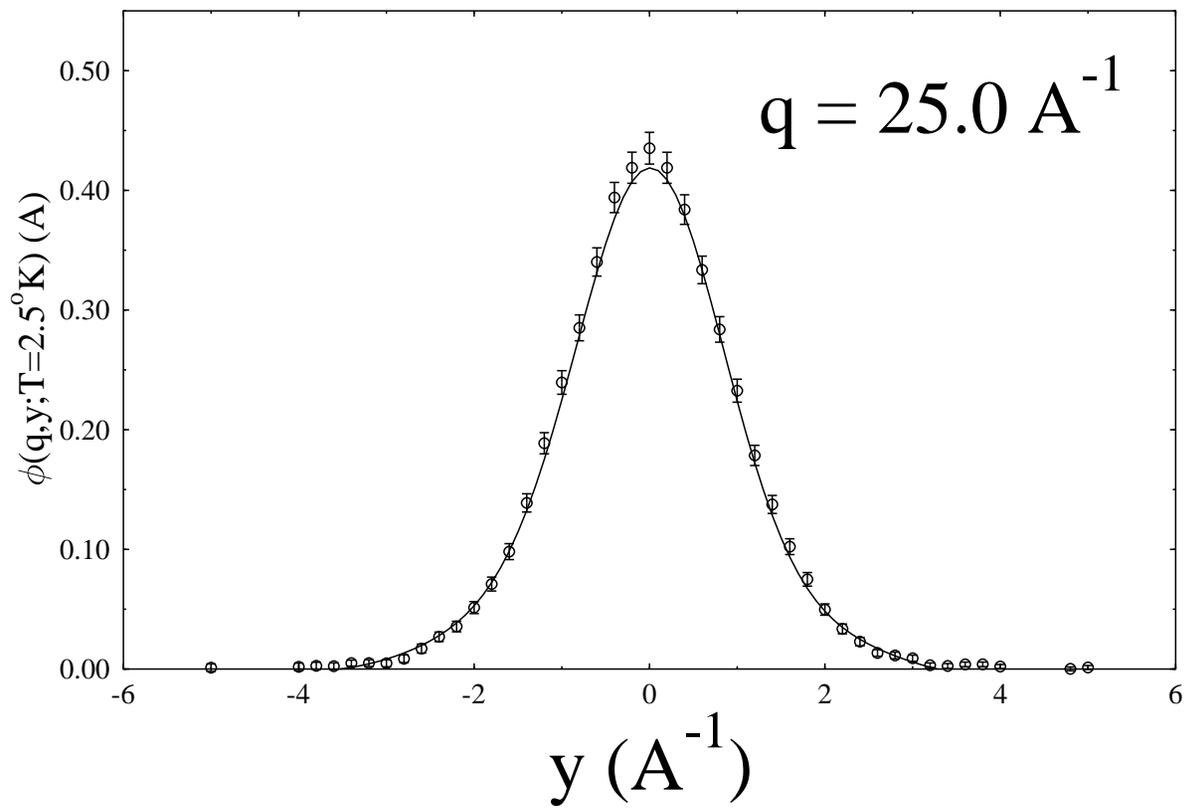
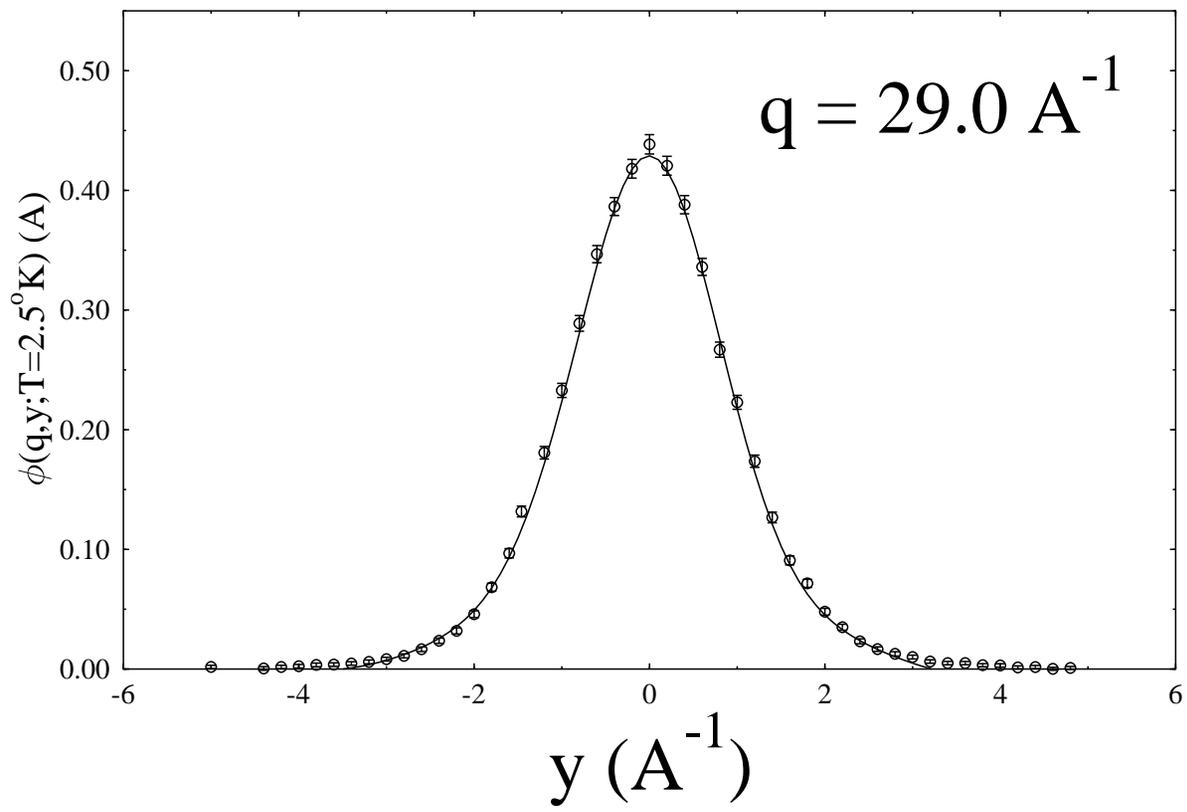
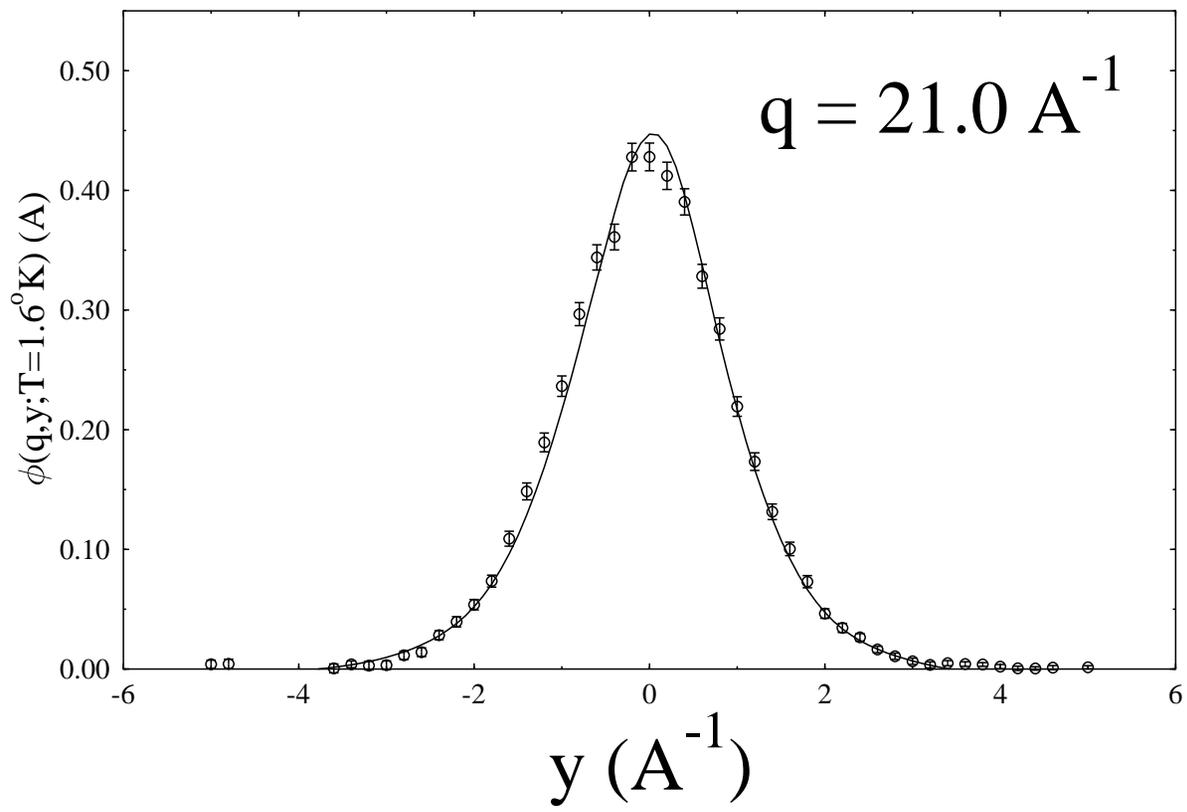
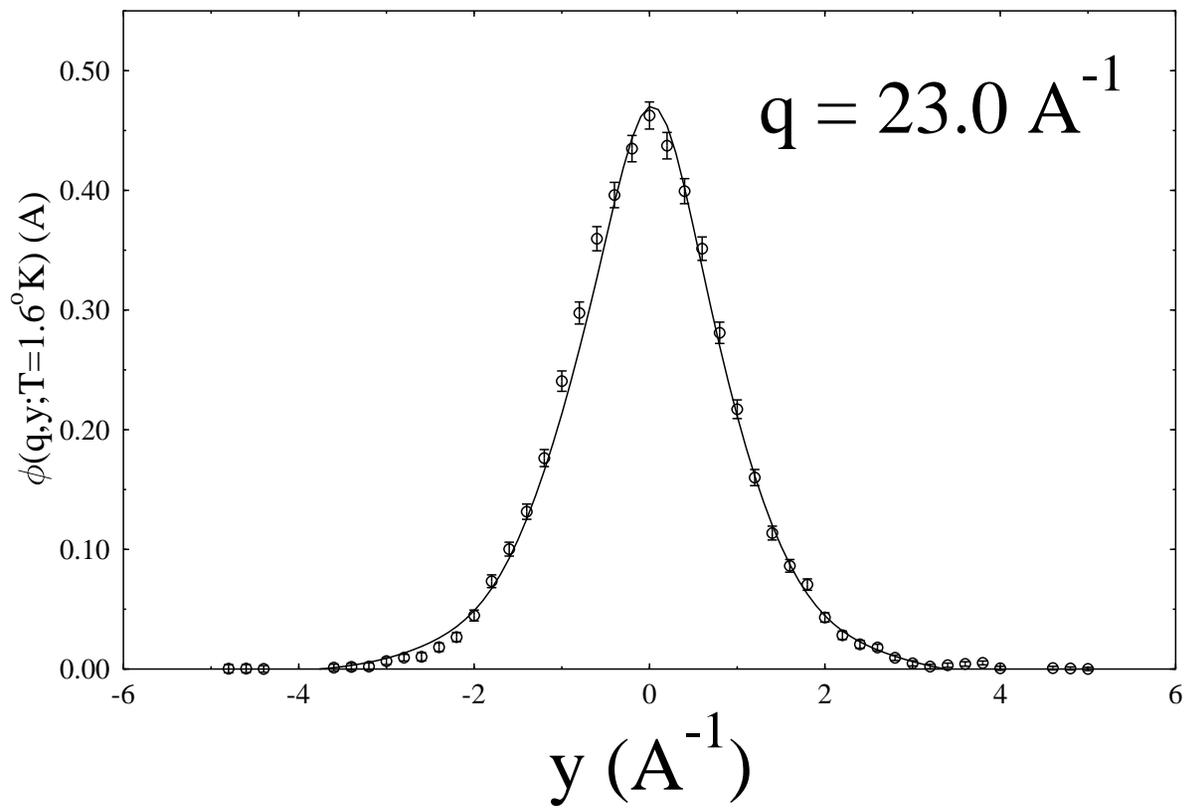
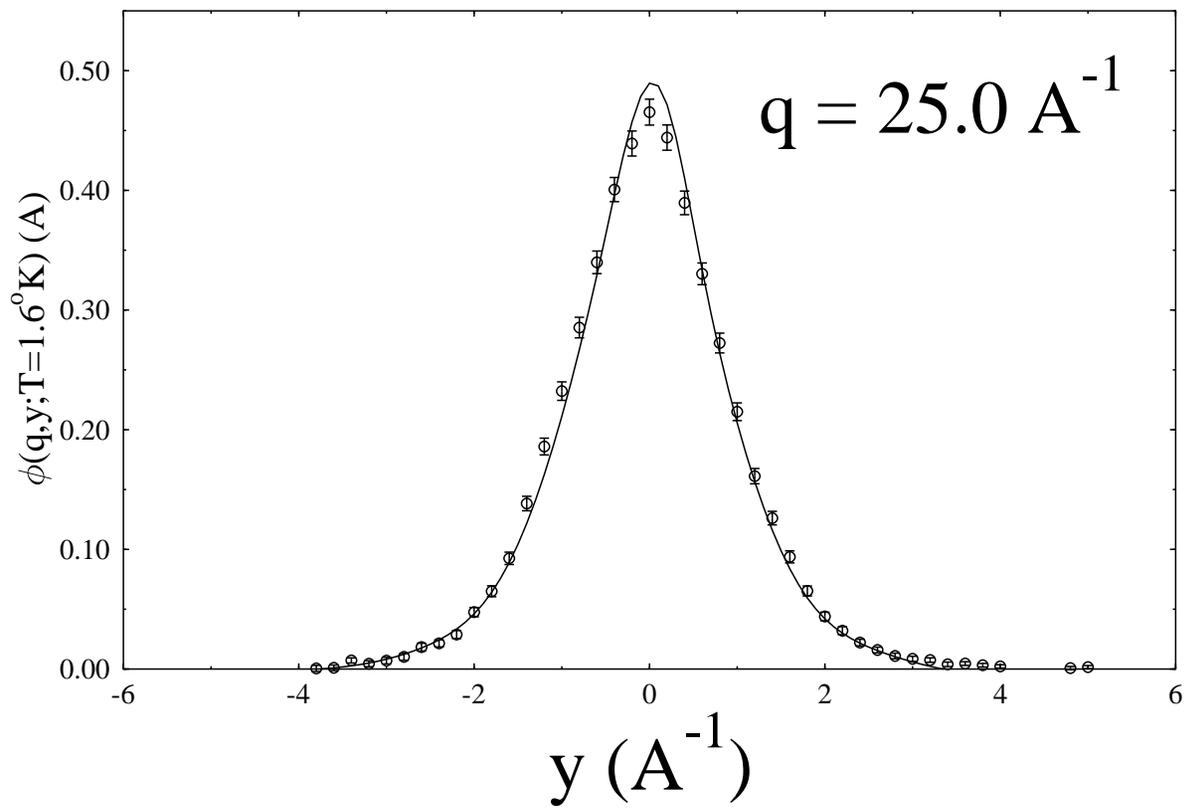
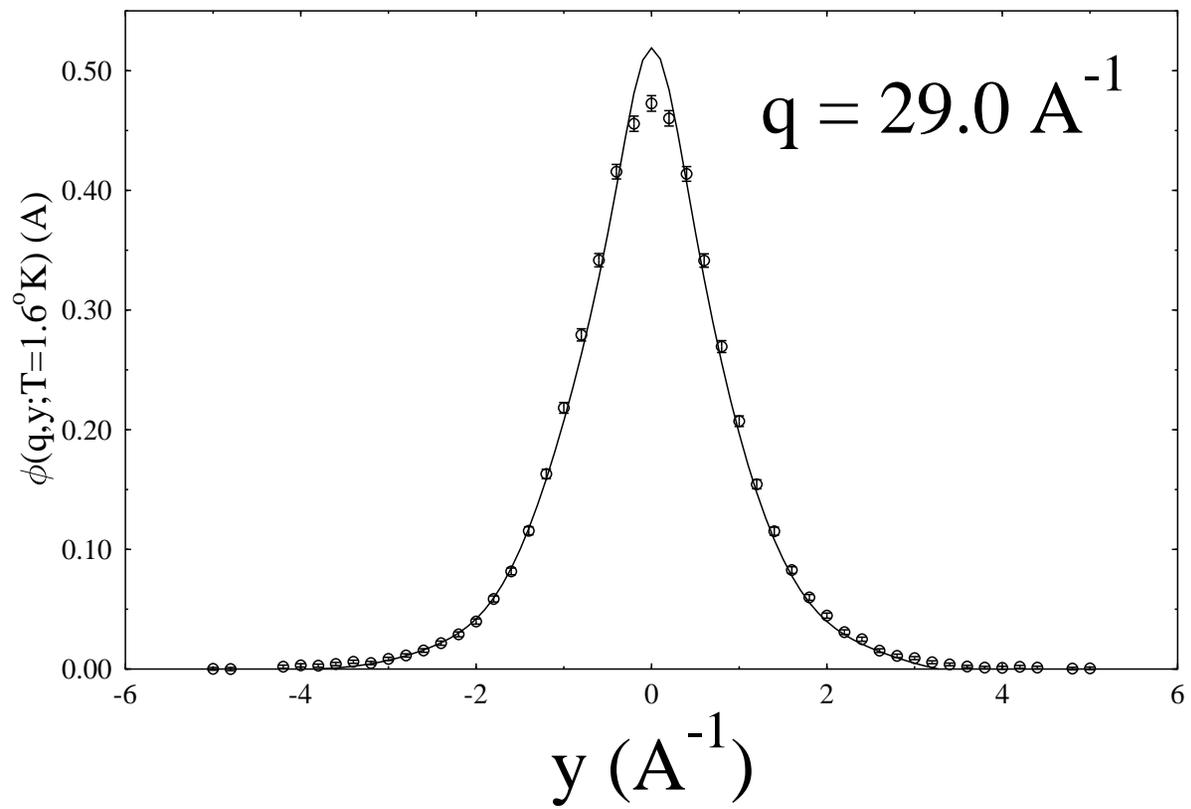
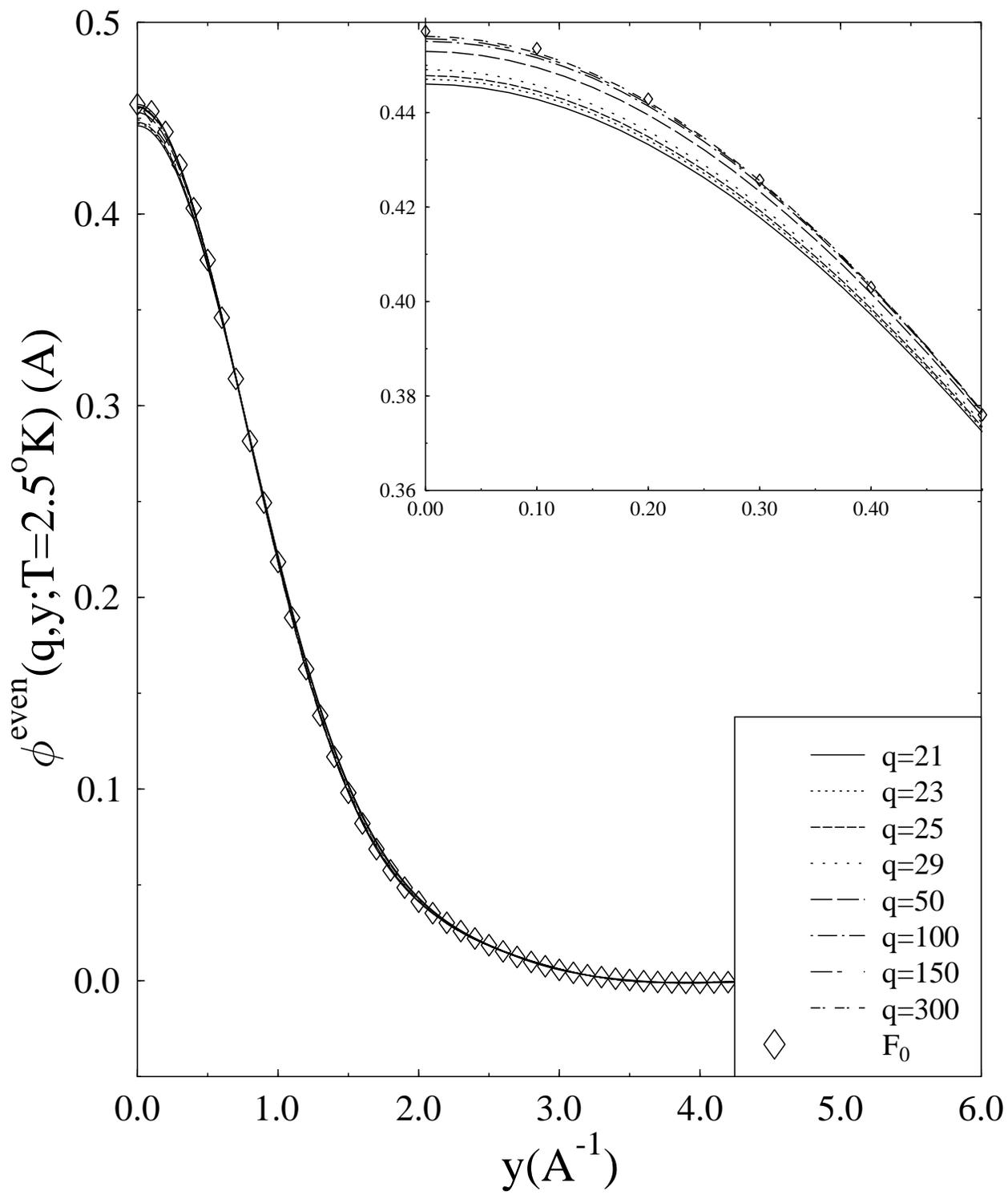
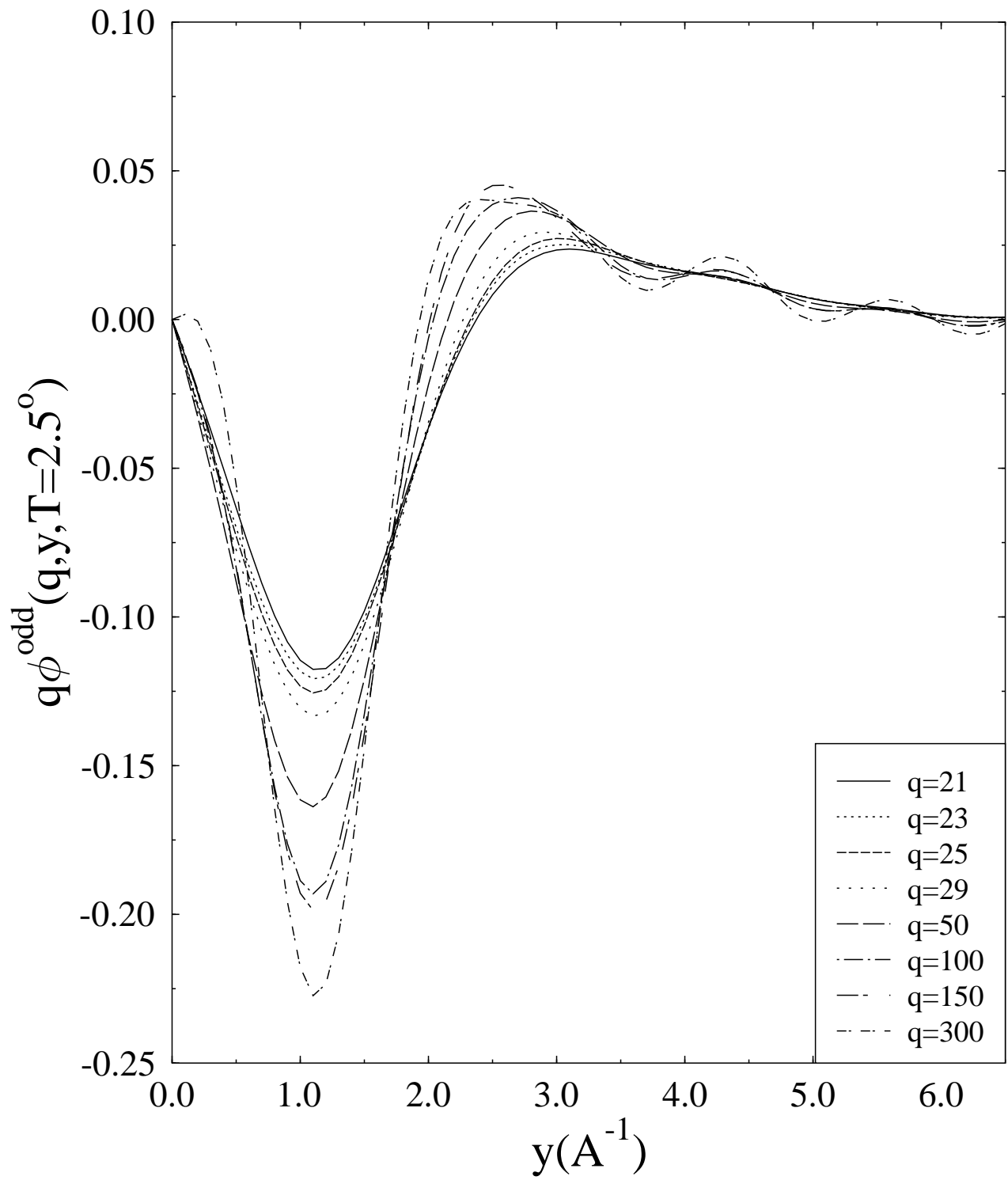
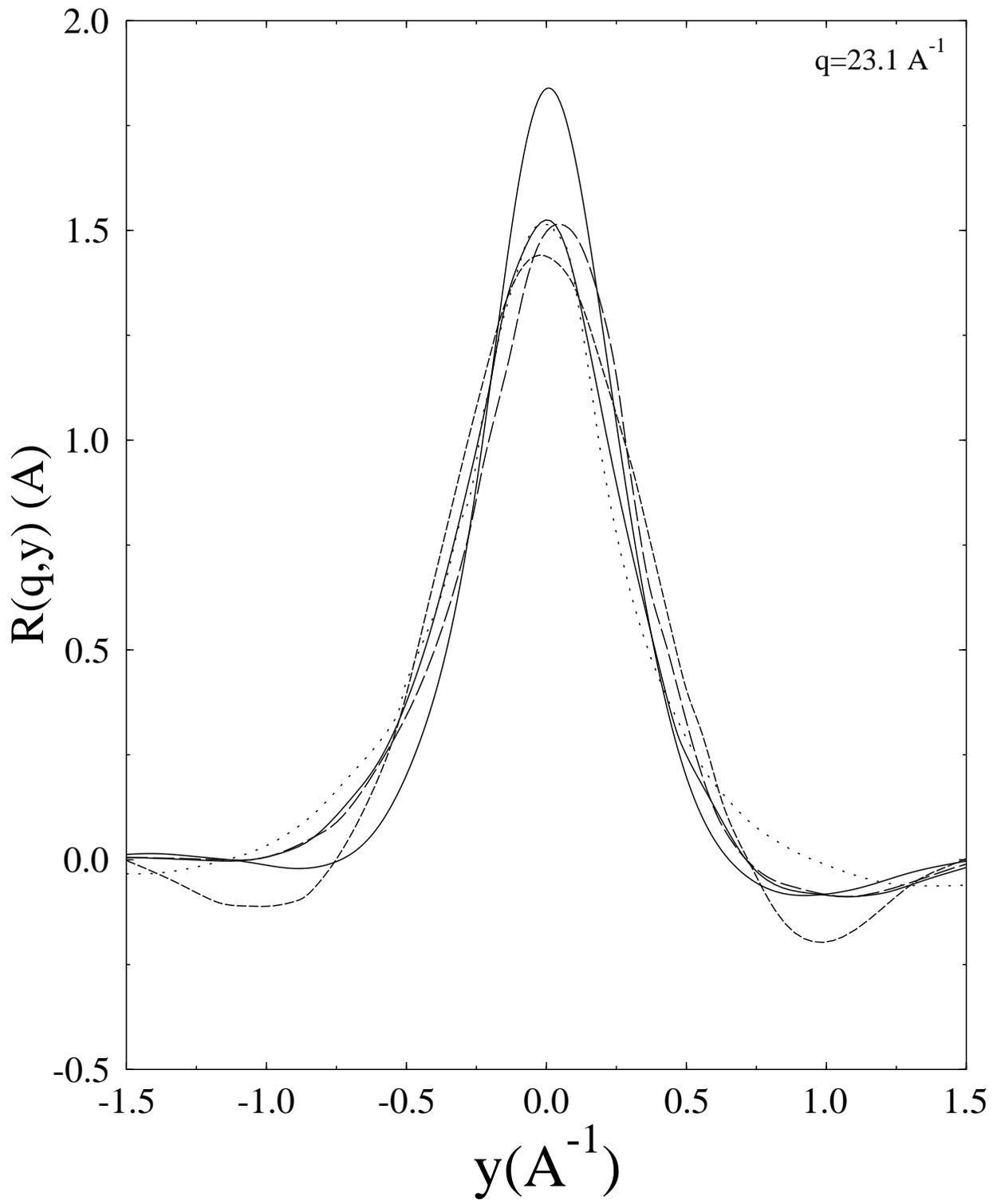
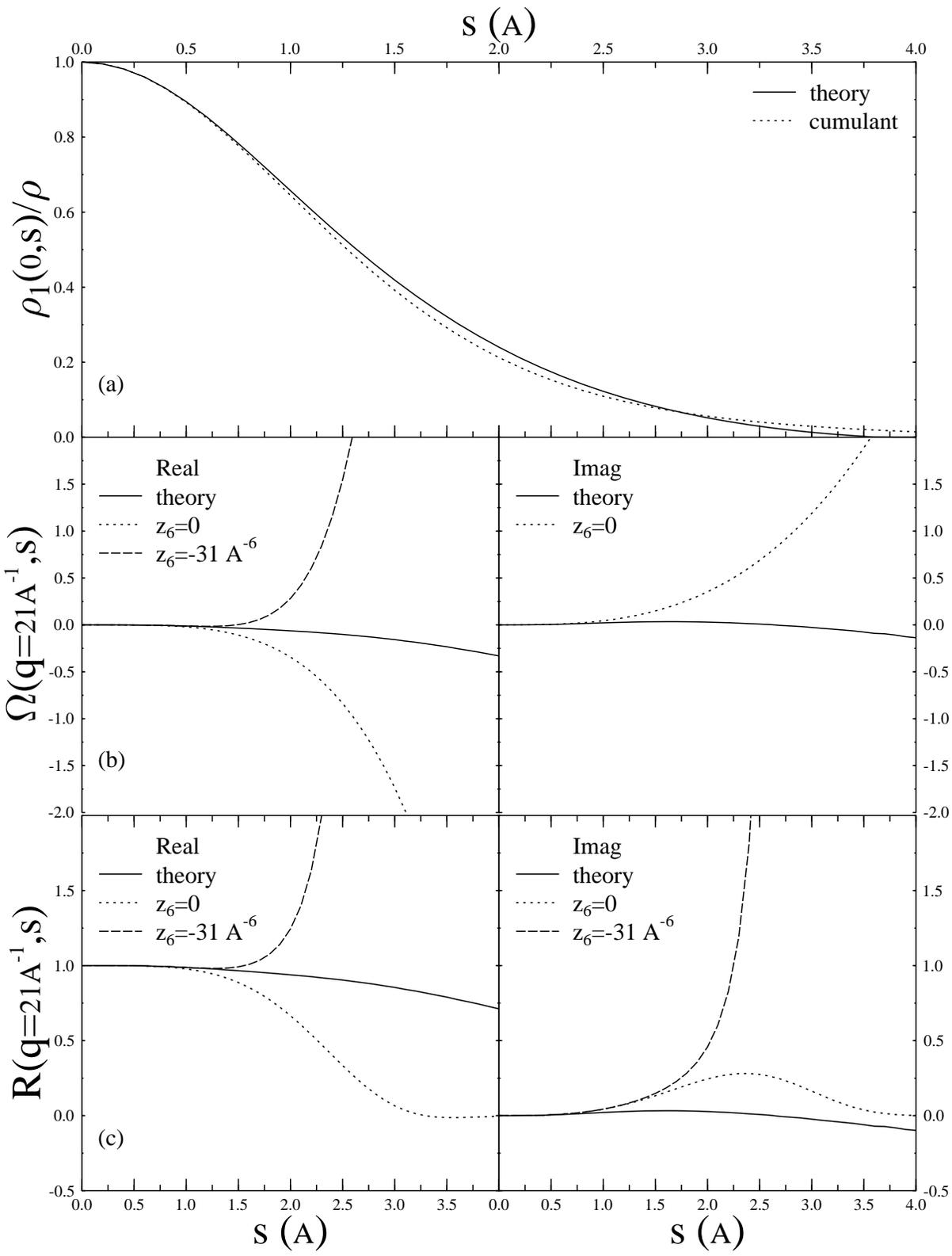
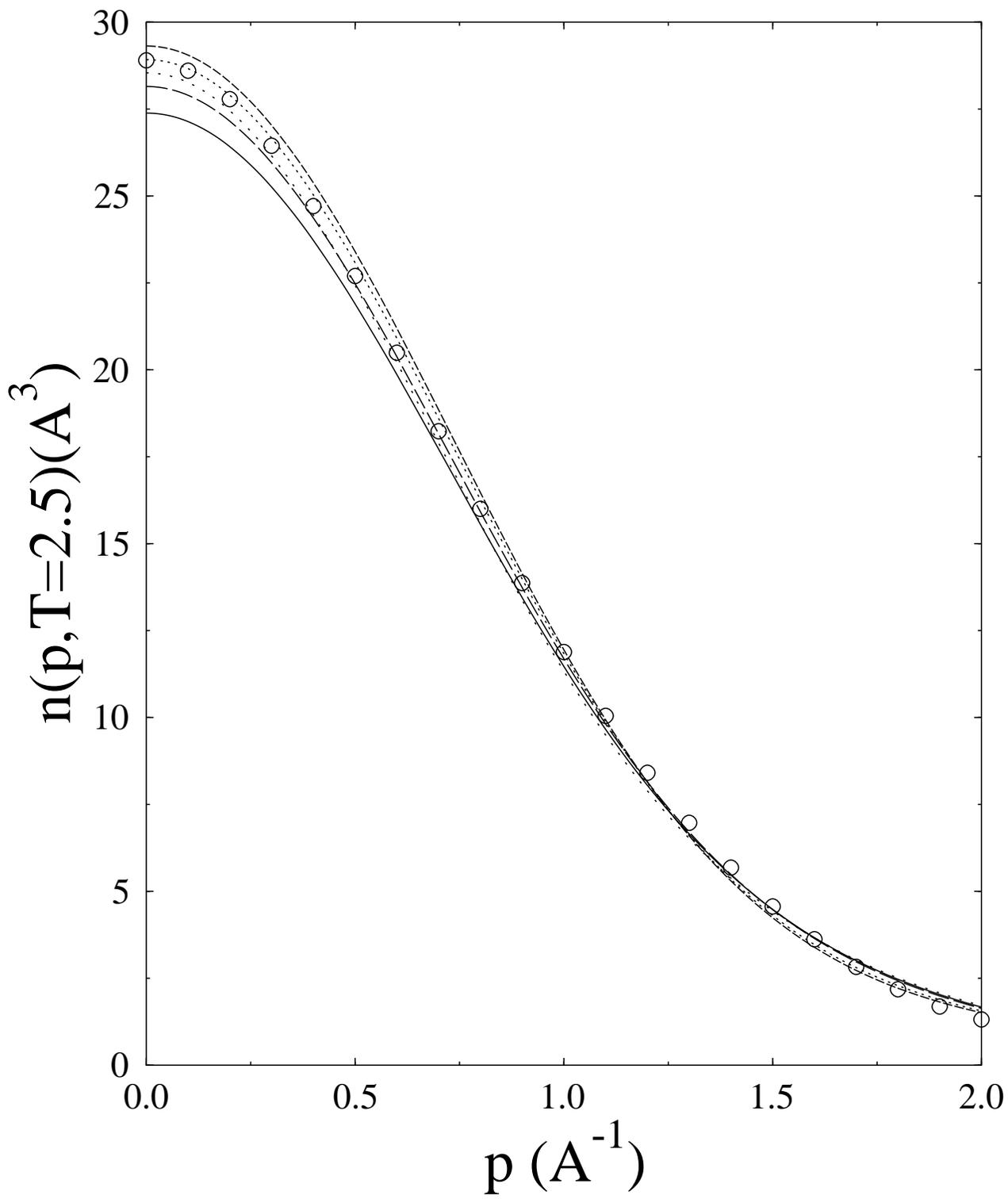
\end{document}